\documentclass[sigconf]{acmart} 
\AtBeginDocument{
  }
\copyrightyear{2025}
\acmYear{2025}
\setcopyright{cc}
\setcctype{by}
\acmConference[CHI '25]{CHI Conference on Human Factors in Computing Systems}{April 26-May 1, 2025}{Yokohama, Japan}
\acmBooktitle{CHI Conference on Human Factors in Computing Systems (CHI '25), April 26-May 1, 2025, Yokohama, Japan}\acmDOI{10.1145/3706598.3714205}
\acmISBN{979-8-4007-1394-1/25/04}

\usepackage{enumitem}
\usepackage{longtable}
\usepackage{booktabs}
\usepackage{tabularx}
\usepackage{xcolor} 
\usepackage{soul}  
\newcommand{\aaa}[1]{\textcolor{black}{#1}}

\AtBeginDocument{
  }

\acmConference[CHI '25]{The ACM CHI Conference on Human Factors in Computing Systems}{April 26--May 1, 2025}{Yokohama, Japan}

\begin{document}

\title{How Users Who are Blind or Low Vision Play Mobile Games: Perceptions, Challenges, and Strategies}

\author{Zihe Ran}
\affiliation{
  \institution{Communication University of China}
  \city{Beijing}
  \country{China}
}
\email{rougetinge@Gmail.com}

\author{Xiyu Li}
\affiliation{
  \institution{Communication University of China}
  \city{Beijing}
  \country{China}
}
\email{elviaovo@gmail.com}

\author{Qing Xiao}
\affiliation{
  \institution{Human-Computer Interaction Institute, Carnegie Mellon University}
  \city{Pittsburgh}
  \state{Pennsylvania}
  \country{USA}
}
\email{qingx@andrew.cmu.edu}

\author{Xianzhe Fan}
\affiliation{
  \institution{Tsinghua University}
  \city{Beijing}
  \country{China}
}
\email{fxz21@mails.tsinghua.edu.cn}

\author{Franklin Mingzhe Li}
\affiliation{
  \institution{Human-Computer Interaction Institute, Carnegie Mellon University}
  \city{Pittsburgh}
  \state{Pennsylvania}
  \country{USA}
}
\email{mingzhe2@cs.cmu.edu}

\author{Yanyun Wang}
\affiliation{
  \institution{College of Media, Communication and Information, University of Colorado Boulder}
  \city{Boulder}
  \state{Colorado}
  \country{USA}
}
\email{mia.wang@colorado.edu}

\author{Zhicong Lu}
\affiliation{
  \institution{Department of Computer Science, George Mason University}
  \city{Fairfax}
  \state{Virginia}
  \country{USA}
}
\email{zlu6@gmu.edu}

\begin{abstract}
As blind and low-vision (BLV) players engage more deeply with games, accessibility features have become essential. While some research has explored tools and strategies to enhance game accessibility, the specific experiences of these players with mobile games remain underexamined. This study addresses this gap by investigating how BLV users experience mobile games with varying accessibility levels. Through interviews with 32 experienced BLV mobile players, we explore their perceptions, challenges, and strategies for engaging with mobile games. Our findings reveal that BLV players turn to mobile games to alleviate boredom, achieve a sense of accomplishment, and build social connections, but face barriers depending on the game's accessibility level. We also compare mobile games to other forms of gaming, highlighting the relative advantages of mobile games, such as the inherent accessibility of smartphones. This study contributes to understanding BLV mobile gaming experiences and provides insights for enhancing accessible mobile game design.

\end{abstract}



\maketitle

\section{Introduction}

\aaa{The gaming industry is experiencing rapid growth worldwide, with mobile gaming leading the way. In 2023, mobile games dominated the gaming market, generating 49\% of the total \$90.4 billion in revenue, followed by computer games at 29\%~\cite{muhammad2024mobile}. Considering the limited digital resources and the digital divide, the advantages of mobile games are even more pronounced in the Global South. The number of players in sub-Saharan Africa nearly tripled, from 77 million in 2015 to 186 million in 2021~\cite{Yieke2024}. This growth is primarily due to the rapid spread of mobile gaming, with 95\% of players in sub-Saharan Africa playing on smartphones or tablets, rather than on consoles or personal computers~\cite{Yieke2024}. However, despite the proliferation of mobile games, the experience of BLV players remains largely underexplored.}

As BLV players increasingly engage with video games~\cite{Bunting2023, nair2024surveyor}, there has been a rise in the development of non-visual games (e.g., audio games)~\cite{andrade2019playing, archambault2007computer, urbanek2019unpacking}, alongside the integration of accessibility features in many mainstream games.

Some HCI research has also explored the design of accessible gaming tools, focusing on balancing accessibility for BLV players with the complex experience of gaming~\cite{nair2021navstick, nair2024surveyor, westin2004game, trewin2008powerup, drossos2015accessible}. However, a significant portion of mainstream games still lack sufficient accessibility features. 

Previous studies have shown that BLV individuals are enthusiastic about games with varying levels of accessibility~\cite{gonccalves2023my}, although much of the prior research has focused primarily on their experiences with accessible games aimed at BLV players~\cite{urbanek2019unpacking, bolesnikov2022understanding}.

Mobile games, due to their widespread use and portability, have become an increasingly popular choice among BLV players~\cite{islam2020design, mahardhika2019mobile}. However, despite the dominance of mobile games as a mainstream form of entertainment~\cite{statista2024mobile}, the experiences of BLV players are often overlooked. Some studies have explored game accessibility for BLV players in areas such as computer and tabletop games, but few HCI scholars have examined the mobile gaming experience and its accessibility for BLV users. There is a notable gap in research regarding how mobile game features—such as popularity and portability—are integrated into mobile gaming experiences for BLV players and how they navigate games with different levels of accessibility on mobile platforms. This represents a significant gap in the literature, particularly concerning how BLV players experience mobile games in real-world settings. Given the rising prevalence of mobile games and their unique position within the gaming ecosystem, understanding how BLV players interact with these games is essential for advancing accessible game design.

To explore these problem spaces, especially considering the Global South context with limited accessibility, we conducted semi-structured interviews with 32 experienced BLV mobile game players in China, each having played mobile games for over 200 hours and engaging in gameplay for at least three hours per week. We focus on the following research questions throughout the studies:

\textbf{RQ1:} What are the perceptions of BLV players regarding playing mobile games?

\textbf{RQ2:} What challenges do BLV players face while playing games and how do they experience gaming on mobile platforms?

\textbf{RQ3:} How do different levels of accessibility in mobile games impact the gaming experience of BLV players?

Our research shows that mobile games, \aaa{with their high accessibility, strong portability, spacetime flexibility, and high popularity, }can help BLV players alleviate anxiety, access information, and build connections with others, thereby fostering a sense of achievement and inclusivity—something that is difficult to achieve in the physical world of the Global South. We found that BLV players use mobile games as a resource to compensate for other accessibility limitations \textbf{(RQ1)}. BLV players face challenges in games, including a lack of accessible tools, design limitations, and the need to invest more effort than sighted players. While BLV-specific games offer higher accessibility, their playability is weak, affecting the overall experience. On mobile platforms, these challenges often limit BLV players' engagement \textbf{(RQ2)}.

The accessibility level of mobile games directly impacts BLV players' experience. High-accessibility games enable deep social interaction and convey auditory information well, but have weak playability. Mainstream games with assistive features balance playability and accessibility, but come with higher cognitive costs and technical challenges. Despite strong motivation, mainstream games without accessible features make participation nearly impossible. The ableism of this design causes psychological harm, and only 4 of 32 participants continue to play. However, some players explored alternative strategies, \aaa{and our participants expressed expectations for future mobile game design} \textbf{(RQ3)}.

In summary, our study makes the following contributions to the HCI community: 
\begin{itemize}
\item We investigate the unique case of BLV mobile game players by conducting interviews with 32 experienced BLV mobile players, gaining insights into their perceptions, barriers, and strategies for playing mobile games, \aaa{and present their expectations for future mobile game design.}
\item We examine the factors influencing BLV players' choice of mobile games by comparing the impact of different types of games on their choices.
\item We analyze how different levels of accessibility in mobile games affect the gaming experiences of BLV players, demonstrating their strong motivation to play mobile games and how they navigate or are hindered by the varying degrees of accessibility available within these mobile games.
\end{itemize}

\section{\aaa{Background and }Related Work}
Our research builds on prior academic studies on the meaning of gaming, mobile accessibility, and game accessibility. By combining insights from these areas, we examined how the accessibility of mobile games affects the gaming experiences of BLV users.

\subsection{Gaming and BLV Individuals}
\aaa{The HCI community has increasingly focused on accessible gaming \cite{holloway2019disability, larreina2024audio}. Studies show that players with disabilities have similar gaming desires as players without disabilities, including enjoying fun, overcoming challenges, and participating in competition.\cite{beeston2018characteristics,waddington2015participatory}}

Previous Research revealed that BLV individuals often experience reduced well-being and increased feelings of loneliness~\cite{karlsson1998self}. Nair et al. found that games can serve as a medium to alleviate feelings of loneliness \cite{nair2024surveyor}. \aaa{Cairns et al. explored the motivations of players with disabilities, identifying social connection, leisure, escapism, and enjoyment as key reasons for engaging with games \cite{cairns2021enabled}. }Video games can promote social interaction, enhance emotional well-being, and increase feelings of social closeness~\cite{granic2014benefits}. Aguado-Delgado and colleagues further suggested that video games might improve the quality of life for BLV individuals, but this potential is hindered by the lack of accessible mainstream games~\cite{aguado2020accessibility}. Ensuring widespread, easy access to mainstream games for the BLV community remains a significant challenge. \aaa{As a common form of gaming, tabletop games often rely on visual elements for communication, posing accessibility barriers for BLV players \cite{andrade2019playing, da2019let}. Similarly, modern digital games frequently emphasize 3D concepts and visual elements, which are often incompatible with assistive technologies, creating challenges for BLV players during gameplay \cite{andrade2020introducing, yuan2011game, bolesnikov2022understanding}.}

\aaa{
Currently, there is a growing body of HCI research aimed at enabling BLV individuals to play games \cite{archambault2008towards, archambault2005make, heron2018eighteen, von2006improving, westin2011advances}, allowing them to experience more advanced computer gaming technologies and enjoyment. At the same time, industry efforts have been made to enhance computer accessibility, such as Microsoft's development of Microsoft Active Accessibility and the W3C's Web Accessibility Initiative. However, these accessibility features have proven to be effective for desktop applications, but remain challenging to implement effectively for computer games \cite{archambault2007computer}. Andrade et al. highlighted in their study of BLV gaming experiences that PC gaming platforms (e.g., Steam) are often difficult to navigate due to insufficient accessibility features \cite{andrade2019playing}.
}

\aaa{
In the realms of console and VR gaming, accessibility has gradually gained attention. Beyond optimizing game design, hardware innovations have opened new possibilities for accessible gaming experiences. For example, Microsoft introduced the Xbox Adaptive Controller in 2019 \cite{xbox_accessibility}, incorporating diverse feedback mechanisms such as speech-to-text, voice commands, and narration to help BLV players better engage with games. Several HCI studies have also developed more inclusive spatial games for BLV players \cite{kamath2024playing,chai2019hungry}. While these advancements provide a more inclusive gaming environment, challenges persist—particularly the reliance on visual and 3D concepts in console and VR game design, which continues to hinder the realization of fully accessible experiences.}

\subsection{Mobile Accessibility for BLV Individuals}
With the rapid development of mobile devices and the widespread availability of the internet, an increasing number of BLV individuals now have access to smartphones~\cite{abraham2022smartphone, tan2024exploration,li2017braillesketch}. These advancements have empowered BLV users to connect to the digital world in ways that were previously unimaginable, allowing them to perform everyday tasks, such as communication, navigation, and information retrieval, more independently~\cite{pundlik2023impact,li2024recipe,li2021non}. Mobile devices, with their built-in accessibility features like screen readers, voice assistants, and haptic feedback, have become essential tools for BLV users, offering new opportunities for engagement, social interaction, and entertainment~\cite{tan2024exploration,tan2023training,li2022feels,kianpisheh2019face}. As a result, mobile technology has become a vital platform not only for improving accessibility but also for enhancing the overall quality of life for the visually impaired community~\cite{della2017teenagers,li2024contextual}. Previous studies have shown that BLV users have a stronger preference for using the mobile version of Facebook over the desktop version~\cite{ramayah2015understanding}. They are eager to use mobile devices to integrate into society and connect with their communities~\cite{wu2014visually,della2017teenagers, grober2016analysis}. \aaa{The accessibility of entertainment on mobile devices is crucial to the quality of life for BLV individuals \cite{baumgartner2023if, jun2021exploring, grober2016analysis, branham2015invisible, li2023understanding, wu2014visually, della2017teenagers}. BLV users actively engage in online entertainment activities such as live streaming to achieve social inclusion \cite{jun2021exploring, rong2022feels}. Additionally, they rely on mobile devices for online shopping and socializing, helping to reduce loneliness and overcome barriers commonly encountered in offline environments \cite{alluqmani2023barriers}.}

In many regions of the Southern Hemisphere, mobile phones are often the only accessible or affordable technology for a significant portion of the population, including those with visual impairments. Due to the high cost, limited availability, and steep learning curve associated with other devices, mobile phones, with their relatively affordable price, wide availability, and user-friendly interface, have become an important alternative assistive device for the BLV community.~\cite{abraham2022smartphone}. Notably, 95\% of BLV internet users in China are proficient in operating mobile phones~\cite{CNNIC2021}.  

However, the current accessible features on mobile devices have limitations, such as the lack of advanced screen readers or limited haptic feedback. These limitations restrict the user experience, especially when performing complex operations and interactions, highlighting the urgency of addressing accessibility issues~\cite{rodrigues2020open}. Furthermore, the availability of accessible apps and services in local languages is often restricted, exacerbating the digital divide and making it more difficult for Global South BLV users to fully participate in the digital world~\cite{Batagoda2024}. Most touchscreens lack audio or tactile feedback, making it difficult for BLV users to locate items on the screen, requiring additional time and effort to search for information \cite{branham2015invisible}. Sometimes, they may even search for things that do not exist, a phenomenon scholars refer to as "not knowing what you don’t know" \cite{bigham2017effects}. Research indicates that the main challenges BLV users face when using the internet include slow page loading, lack of alternative formats, complex information architecture, and broken links \cite{bigham2017effects}. 

This issue is even more pronounced in the Global South. Visual impairment disproportionately affects populations in low and middle-income countries (LMICs), where 90\% of those with visual disabilities are found~\cite{bourne2017magnitude, abraham2022smartphone}. In addition, the challenges related to accessibility are compounded by the widespread reliance on mobile devices as the primary means of accessing information and services in Global South~\cite{correa2016digital}. Therefore, improving the accessibility of mobile devices, especially in the entertainment and social domains, has become an urgent issue to address in these regions. 

\aaa{In recent years, mobile games have dominated the global gaming market, holding the largest market share. In 2024, the global mobile gaming market is expected to generate \$98.74 billion in revenue. According to Statista, the market size of mobile games is expected to continue growing, reaching \$118 billion by 2027 \cite{statista2024mobile}. However, despite the attention given to the accessibility of live streaming and social platforms for BLV users, the accessibility of mobile games remains underexplore in the literature. As the mobile gaming market continues to grow, it is becoming increasingly important to enhance its accessibility for BLV users.}

\subsection{Accessibility Gaming Tools for BLV Individuals}
Video games have increasingly become a significant medium through which people perceive and interact with the world, serving as an essential means to enhance user well-being and foster social connections~\cite{ballou2024platform, vuorre2023affective, larreina2024audio}. However, the strong visual focus of these games and their lack of accessibility result in a significantly reduced or deprived experience for a large number of users, especially those with visual impairments \cite{metatla2020robots}. Previous scholars have argued that enhancing the accessibility of video games, enabling BLV users to have comparable experiences, is both a moral and legal obligation in today's information society~\cite{sekhavat2022sonification}.

\aaa{Some HCI scholars are working to develop innovative design strategies and technological solutions that make games more inclusive, such as incorporating audio cues, haptic feedback, and screen reader compatibility to replace or supplement visual information ~\cite{islam2020design, dobosz2016control, milne2014brailleplay, martinez2024playing,atkinson2006making,yuan2008blind}. }Coronado et al.~\cite{coronado2023game} proposed 35 guidelines for adapting games for BLV individuals, including meaningful and clear audio feedback, increased co-design with BLV users, among others. Despite efforts by many game companies, designers, and researchers to provide BLV users with gaming experiences equivalent to those of sighted players, achieving this goal remains highly challenging in practice~\cite{martinez2024playing, smith2018rad, chai2019hungry,milne2014brailleplay, branham2015collaborative, bolesnikov2022understanding}. Most games accessible to BLV players often present an overwhelming amount of information, requiring additional cognitive effort to discern and filter useful content, which significantly impacts the fluidity and speed of their gameplay experience~\cite{branham2015collaborative, bolesnikov2022understanding}. In some cases, BLV players can access simplified versions of games designed for sighted users; however, their experience is typically limited to following game instructions, making equitable access nearly impossible~\cite{agrimi2024game, gonccalves2023my, bolesnikov2022understanding,smith2018rad}. 

Additionally, adapting existing games to accommodate the BLV community poses a conflict between maintaining the game’s complexity and preserving its pacing. Researchers have examined various levels of accessibility in games to better address the needs of BLV individuals~\cite{zhang2024designing, stadler2018blind, gonccalves2023my, bolesnikov2022understanding, coronado2023game, rodriguez2018cprforblind, giannakopoulos2018accessible, gluck2021racing, rodriguez2018arduino}. These efforts include studying mainstream games designed for sighted players~\cite{gonccalves2023my}, developing inclusive games specifically tailored for visually impaired users~\cite{stadler2018blind}, and creating assistive plugins to enhance the accessibility of mainstream games~\cite{nair2024surveyor}.

Therefore, games can be categorized into the following three types based on their level of accessibility:
\begin{itemize}
    \item \textbf{Games Without Accessibility Tools.} These are mainstream games primarily designed for sighted players, with no built-in accessibility features or external accessible tools~\cite{gonccalves2023my}. Some BLV players still attempt to play these games by relying on spatial and audio cues to perceive the position and direction of elements~\cite{gonccalves2023my}. For example, players use background music and sound effects within the game to understand their current situation. Although games like \textit{Pokémon} were not designed with visually impaired individuals in mind, players can make them more accessible by relying on actions such as bumping into objects and using 3D audio effects~\cite{kovac2018pokemon}.
    \item \textbf{Games with Accessibility Tools.} These games are mainstream titles with some built-in accessibility features or have assistive plugins designed specifically for the game~\cite{bolesnikov2022understanding, nair2024surveyor, gonccalves2023my}. BLV players can use these tools to navigate and play the game. For instance, in \textit{Hearthstone}, a mod named "Hearthstone Access" was developed to assist BLV players by reading out screen content and enabling operations such as card collection, sorting, and crafting within the game~\cite{hearthstoneaccess2024}.
    \item \aaa{\textbf{BLV-Specific Mobile Games. }These games typically use audio information (such as screen readers and text-to-speech technologies) to convey visual details. \aaa{These games are fully accessible for BLV players, with features like gesture systems and visual focus systems, allowing seamless play without the need for additional accessible tools~\cite{stadler2018blind, westin2004game}.} Classic text-based adventure MUD games or audio games have low development costs and relatively simple gameplay \cite{Audio2020,smith2018rad,drossos2015accessible,bolesnikov2022understanding,kirke2018soundtrack}; their high accessibility is achieved through built-in screen readers and audio design provided by the game itself~\cite{westin2004game,andresen2002playing}.}
\end{itemize}

Giannakopoulos et al. identified three types of barriers in games~\cite{giannakopoulos2018accessible}: Mainstream games, although playable, are not optimized for BLV players; most video games and audio games are either too simplistic or too outdated to provide engaging experiences and may lack intuitive game introductions; and only a small number of video games can be accessed through accessibility tools. However, their work merely touched on these issues without delving into how mobile game accessibility specifically shapes the experiences of BLV users. \aaa{While most research focuses on Western contexts, our work highlights the unique experiences of Chinese BLV community in mobile gaming.} Furthermore, no research has compared how mobile games with varying levels of accessibility meet the needs and preferences of BLV users. Building on previous literature, we explored how BLV users in China experience mobile games with different levels of accessibility.

\section{Method: Semi-Structured Interviews}

We conducted one-on-one semi-structured interviews with 32 blind and low vision (BLV) mobile game players (P1 to P32) to gain an understanding of their daily lives and explore their experiences across different types of games. Each interview lasted around two hours, with six conducted in person and 26 conducted online. By focusing on their experiences with different game types, we sought to uncover their perceptions of playing mobile games specifically, as well as the barriers they face, such as accessibility challenges, game design limitations, and social factors that influence their gaming habits. 

We also \aaa{focus on} the specific challenges and obstacles these BLV players encounter while engaging with mobile games, including issues related to game mechanics, navigation, and the lack of inclusive design features. We studied how these BLV players manage and navigate different levels of accessibility within mobile games, from those with full accessibility features to those with minimal or no accommodations for BLV users. Our goal was to understand the strategies they employ to adapt to these varying levels of accessibility, the impact on their overall gaming experience.

\begin{table*}[h] 
  \caption{Information of participants. The corresponding game type numbers are as follows: 1. Games Without Accessibility Tools 2. Games with Accessibility Tools 3. BLV-Specific Mobile Games }
  \label{tab:participants} \label{app:participants}
  \centering
  \small
  \setlength{\tabcolsep}{5pt} % Adjusts the column spacing
  \begin{tabular}{>{\centering\arraybackslash}p{0.5cm} >{\centering\arraybackslash}p{0.5cm} >{\centering\arraybackslash}p{1.8cm} >{\centering\arraybackslash}p{3.5cm} >{\centering\arraybackslash}p{2cm} >{\centering\arraybackslash}p{2.5cm} >{\centering\arraybackslash}p{2cm}}
    \hline
    \textbf{ID} & \textbf{Age} & \textbf{Gender} & \textbf{Occupation} & \textbf{Vision Status} & \textbf{Daily Playtime} & \textbf{Game Types} \\
    \hline
    P1  & 30 & Male   & Game Designer & Blind & 4 hours & 2, 3 \\
    P2  & 24 & Non-binary   & Unemployed  & Blind & 1 hour & 1, 3 \\
    P3  & 28 & Male   & Massage Therapist & Blind & 3 hours & 3 \\
    P4  & 25 & Female & Sales & Low Vision & 3 hours   & 1, 2, 3 \\
    P5  & 20 & Male & Undergraduate & Low Vision  & 3 hours & 1, 2, 3 \\
    P6  & 27 & Male   & Doctor  & Low Vision   & 1 hour & 1, 2, 3 \\
    P7  & 21 & Male   & Game Tester & Blind & 5 hours    & 1, 2, 3 \\
    P8  & 45 & Male   & Novelist  & Blind & 1 hour & 1, 3 \\
    P9  & 18 & Female & High School Student  & Blind & 30 min & 1, 2, 3 \\
    P10 & 28 & Male   & Unemployed  & Low Vision  & 30 min & 1, 2, 3 \\
    P11 & 23 & Male   & Unemployed   & Blind & 40 min  & 1, 2, 3 \\
    P12 & 19 & Female & High School Student  & Blind & 1 hours & 1, 2, 3 \\
    P13 & 19 & Male   & Unemployed & Blind  & 3 hours  & 1, 2 ,3 \\
    P14 & 35 & Male   & Software Tester  & Blind & 3 hours & 1, 3 \\
    P15 & 25 & Male   & Unemployed & Blind  & 2 hours & 1, 2, 3 \\
    P16 & 21 & Male   & Massage Therapist & Blind& 2 hours & 1, 2, 3 \\
    P17 & 28 & Female & Graduate Student & Blind  & 1 hours& 1, 3 \\
    P18 & 35 & Female & New Media Operator & Blind  & 1 hour & 3 \\
    P19 & 26 & Female & Clothing Store Owner & Blind & 5 hours & 1, 2, 3 \\
    P20 & 35 & Male   & Teacher & Low Vision & 30 min & 2, 3 \\
    P21 & 25 & Male   & Corporate Executive & Blind & 50 min & 2, 3 \\
    P22 & 21 & Male   & Massage Therapist & Blind  & 1 hour & 1, 2, 3 \\
    P23 & 37 & Male   & BLV Board Game Designer  & Blind & 3 hours  & 1, 2, 3 \\
    P24 & 27 & Male   & Online Game Streamer & Blind & 10 hours & 1, 2, 3 \\
    P25 & 19 & Male   & High School Student & Blind  & 4 hours & 2, 3 \\
    P26 & 29 & Female & Online Shop Owner & Blind & 1 hour & 1, 3 \\
    P27 & 34 & Male   & Instrument Tuner & Low Vision & 30 min & 2, 3 \\
    P28 & 26 & Male   & Audiobook Editor & Blind    & 2 hours & 3 \\
    P29 & 20 & Male   & High School Student & Blind & 1 hour  & 1, 2, 3 \\
    P30 & 21 & Male   & Undergraduate   &  Blind  & 20 min & 1, 2, 3 \\
    P31 & 19 & Male   & High School Student  & Low Vision & 3 hours & 2, 3 \\
    P32 & 33 & Female & Online Novel Narrator & Blind & 1 hour & 1, 2, 3 \\
    \hline
  \end{tabular}
\end{table*}

\subsection{Participants: BLV Mobile Players (N=32)}

We recruited 32 BLV mobile players participants through major BLV volunteer organizations in China, online BLV players communities, and specialized online forums and social media groups about gaming (e.g., \textit{Tiantanduping Forum}, \textit{Zhengdu Forum}, \textit{aimang.net}). To participate in our study, participants must be 18 years or older, legally or totally blind, and have had prior game-play experiences with at least 3 hours per week in past one year. Our participant group comprised 23 male players, 8 female players, and 1 nonbinary player, including 26 individuals who are completely blind and 6 with low vision. Participants were aged between 18 and 45 years (M = 25.75, SD = 6.13). \aaa{The interviewed BLV players were not directly compensated for their time; however, as per their preference, we purchased gifts for them. Based on cost differences such as travel expenses, the prices of the gifts ranged from CNY ¥100 to ¥150.} \aaa{The recruitment and study procedure was approved by the Institutional Review Board (IRB).}

Notably, among the BLV players who participated in our study, 6 were unemployed, \aaa{while 26 held various jobs, }including 4 who were actively working as game designers. This diversity in employment status not only reflects the varied socio-economic backgrounds of our participants but also enriches our research perspective. Including game designers, in particular, provides valuable insights into the experiences of BLV individuals who are both users and creators of accessible gaming content. Detailed demographic information could be found in Table~\ref{tab:participants}. 

~\subsection{Study Procedure}
During the semi-structured interviews, we began with demographic questions about their age, gender, vision condition, and occupation (5 minutes). Following this, we prepared a warm-up questions, encouraging participants to share insights about their mobile experience, school experiences, career paths, family, and any challenges they face in their everyday lives (25 minutes). Next, we inquire about their overall opinion with games, experiences with different types of games, with a specific focus on how vision conditions (including total blindness and low vision) affected gaming experiences (30 minutes).

We then asked detailed questions about their view of mobile games, each mobile game they had played and explored their specific experiences and feelings within these games, with a particular emphasis on understanding how vision conditions influenced their gaming experiences (40 minutes). Lastly, we asked about their expectations for mobile games and future game design (20 minutes).

~\subsection{Data Analysis}
In our data analysis, we employed a thematic analysis approach~\cite{kuo2023understanding, braun2019reflecting}. This method was chosen for its flexibility and depth, enabling us to identify, analyze, and report patterns within the data while also considering the broader context and the participants' lived experiences. The reflexive nature of this approach emphasized the active role of the researchers in constructing meanings from the data, acknowledging that our interpretations are shaped by our own perspectives and the research context.

Three researchers collaboratively analyzed approximately 65 hours of recorded interview data. The data was subjected to an open coding process, where the three researchers together reviewed the transcripts to identify and label meaningful segments of text. This process was iterative and involved multiple rounds of coding to ensure that all relevant aspects of the data were captured. We generated a total of 876 codes. Throughout the coding process, the three researchers engaged in continuous discussions to explore and negotiate any potential points of agreement or divergence in their interpretations. These discussions were crucial for ensuring the reliability and validity of our coding framework, as they allowed us to critically examine each other's perspectives, challenge assumptions, and refine our understanding of the data. By fostering an open and reflexive dialogue, we could build a shared understanding of the emerging themes, ensuring that the final codes reflected a comprehensive interpretation of the participants' experiences and insights.

After completing the coding process, we moved on to the next phase of our analysis, which involved conceptualizing higher-level themes from the 876 initial codes. To achieve this, we employed affinity diagramming~\cite{kuo2023understanding} to organize and synthesize our qualitative data by grouping related codes into meaningful clusters. \aaa{ In response to our research questions, we identified three key themes: the external sensory phenomenon and entertainment needs, social depth and information dissemination effectiveness, and the relationship between playability and accessibility. Each theme was further divided into subcategories, such as alleviating boredom, achieving a sense of accomplishment, alternative strategies, and design expectations. In ~\autoref{Findings:Perceptions} and ~\autoref{Findings:Accessibility}, we present the themes derived from the data analysis.} \ref{Schematic Representation Of Findings}
\begin{table*}[]
\caption[Alt text]{Schematic Representation Of Findings. This table depicts the perceptions of BLV users regarding mobile gaming and analyzes the impact of varying levels of accessibility.}
\label{Schematic Representation Of Findings}
\begin{tabular}{lll}
\cline{2-3}
\multicolumn{1}{l|}{} & \multicolumn{1}{l|}{\textbf{\begin{tabular}[c]{@{}l@{}}Finding1: Perceptions of BLV Users \\ for Playing Mobile Games\end{tabular}}} & \multicolumn{1}{l|}{\textbf{\begin{tabular}[c]{@{}l@{}}Finding2: The Impact of Varying Levels of \\ Accessibility in Mobile Games on BLV Players\end{tabular}}} \\ \cline{2-3} 
\multicolumn{1}{l|}{} & \multicolumn{1}{l|}{\textbf{}}                                                                                                       & \multicolumn{1}{l|}{\textbf{1)BLV Game}}                                                                                                                        \\
\multicolumn{1}{l|}{} & \multicolumn{1}{l|}{\textbf{1)Why Mobile Phone?}}                                                                                    & \multicolumn{1}{l|}{- Effective dissemination of auditory information}                                                                                          \\
\multicolumn{1}{l|}{} & \multicolumn{1}{l|}{*External Organ* (N=8)}                                                                                          & \multicolumn{1}{l|}{- Deep social connection in BLV community}                                                                                                  \\
\multicolumn{1}{l|}{} & \multicolumn{1}{l|}{- Popularity}                                                                                                    & \multicolumn{1}{l|}{- High accessibility, low playability}                                                                                                      \\
\multicolumn{1}{l|}{} & \multicolumn{1}{l|}{- Spacetime Flexibility}                                                                                         & \multicolumn{1}{l|}{}                                                                                                                                           \\
\multicolumn{1}{l|}{} & \multicolumn{1}{l|}{- Portability}                                                                                                   & \multicolumn{1}{l|}{\textbf{2)Game with accessibility tools}}                                                                                                   \\
\multicolumn{1}{l|}{} & \multicolumn{1}{l|}{}                                                                                                                & \multicolumn{1}{l|}{- High cognitive cost}                                                                                                                      \\
\multicolumn{1}{l|}{} & \multicolumn{1}{l|}{}                                                                                                                & \multicolumn{1}{l|}{- Technological challenges}                                                                                                                 \\
\multicolumn{1}{l|}{} & \multicolumn{1}{l|}{\textbf{2) Why Mobile Game?}}                                                                                    & \multicolumn{1}{l|}{- Balance between accessibility and playability}                                                                                            \\
\multicolumn{1}{l|}{} & \multicolumn{1}{l|}{*Entertainment Needs* (N=29)}                                                                                    & \multicolumn{1}{l|}{}                                                                                                                                           \\
\multicolumn{1}{l|}{} & \multicolumn{1}{l|}{- Alleviate Boredom}                                                                                             & \multicolumn{1}{l|}{\textbf{3)Mainstram game without accessibility tools}}                                                                                      \\
\multicolumn{1}{l|}{} & \multicolumn{1}{l|}{- A Sense of Achievement Difficult to Attain Elsewhere}                                                          & \multicolumn{1}{l|}{- Strong playing motivation}                                                                                                                \\
\multicolumn{1}{l|}{} & \multicolumn{1}{l|}{- A Pathway for BLV Players to Mainstream Inclusion}                                                             & \multicolumn{1}{l|}{- Psychological barriers caused by ableism in design}                                                                                       \\
\multicolumn{1}{l|}{} & \multicolumn{1}{l|}{- Fostering Deep Social Connections for BLV Users}                                                               & \multicolumn{1}{l|}{- Low accessibility, high playability}                                                                                                      \\
\multicolumn{1}{l|}{} & \multicolumn{1}{l|}{}                                                                                                                & \multicolumn{1}{l|}{- Diverse gameplay strategies and varied design needs}                                                                                      \\ \cline{2-3} 
                      &                                                                                                                                      &                                                                                                                                                                
\end{tabular}
\end{table*}

\section{Findings: Perceptions of BLV Users for Playing Mobile Games} \label{Findings:Perceptions}

\aaa{We first explore why BLV players choose to play mobile games. Our findings indicate that their preference is primarily driven by the widespread accessibility of smartphones in the Chinese market, the ability of mobile games to transcend spatial and temporal boundaries, and their portability. Additionally, mobile games serve as a means to alleviate boredom and achieve a sense of accomplishment that is often hard to find in other aspects of life. Moreover, mobile games provide BLV players with a better way to integrate into mainstream society, fostering social connections by enabling interaction with others and building a sense of community.}

\subsection{\aaa{Why Mobile Phone?}}

\subsubsection{\aaa{Popularity}}
In the mobile era, the majority of respondents (N=32) are proficient in using smartphones, but fewer than half of the BLV players own a computer (N=12). As P27 mentioned, "The operation of computers is very different from smartphones. Almost none of the BLV people I know can use a computer." P30 added, "I spent a long time trying to learn, but I just couldn't  grasp the tricks. However, when I started using a smartphone, I immediately learned because it’s just a touchscreen." Unlike computers, which have a more complex operating system often combined with a keyboard and mouse, smartphones offer simpler touch-based interaction, which BLV users can easily learn.

\aaa{In contrast, the inherent accessibility of smartphones is a significant advantage. Smartphones are equipped with voice reading systems or screen readers, which help overcome various challenges in information retrieval. Games that are compatible with screen readers are prioritized by visually impaired players, as this combination provides a smoother gaming experience, is easier to learn, and aligns well with the habits of visually impaired users. Additionally, P9 noted that smartphones have dedicated gesture systems that can be utilized in games. Beyond touchscreen and button controls, BLV users can use a variety of gestures to provide feedback and input information. The diversity of gesture types and their combinations enables more complex commands, significantly expanding the ways in which games can be controlled.}

\subsubsection{\aaa{Spacetime Flexibility}}
Among our respondents (N=10), computers are seen as being restricted to a particular physical space due to their large size and difficulty in portability. Respondents who have tried console or VR games (N=4) also believe that, apart from handheld consoles, other gaming systems always require some physical space to accommodate multiple components. \aaa{P7, who frequently tests games on different devices, mentioned that the home-use nature of the PlayStation limits him to a specific space, making it the least favored gaming device. In contrast, mobile games require only a smartphone—something that almost every BLV user possesses. }Mobile games also allow for online matching with other players, eliminating the need to rely on the already limited social networks in the real world. Participants mentioned that smartphones are easy to carry, and mobile games are quick to start and end, allowing users to complete all processes independently. P29 explained:
\begin{quote}
\textit{"While waiting for the bus, I can take out my phone and complete a game task. It's quick and simple, allowing me to relieve daily stress during fragmented time. Moreover, mobile games are entirely up to me; I can play whenever I want, without needing to rely on family or friends for assistance with navigation or additional tasks."} (P29)
\end{quote}

\aaa{The convenience of smartphones makes mobile games virtually unrestricted by time and space, reducing additional travel costs and eliminating the need for physical space for gameplay.}

\subsubsection{\aaa{Portability}}
In terms of tabletop games, the lack of accessibility in physical environments forces visually impaired individuals to spend more energy exploring unfamiliar places. P3, P6, P14, and P22 had similar experiences. Despite having various tasks that require them to be physically present, and being accustomed to going out frequently, they still avoid playing tabletop games. P14 mentioned two main reasons: they feel anxious about navigating unfamiliar areas beyond their regular commute, and most tabletop game cafes lack accessibility features.

Particularly, due to physical environmental limitations and limited social networks, they often struggle to find enough players to join in tabletop games. P29, a high school student from a special education school who is keen on exploring new and trendy activities, shared that many of his attempts to organize tabletop games had failed.
\begin{quote}
   \textit{"Several times, we spent a week coordinating everyone’s schedules... Some parents thought it was too troublesome and unnecessary, and others, even with navigation, couldn’t find the exact location. In the end, we completely gave up on these kinds of games."} (P29)
\end{quote}

\aaa{In contrast, the portability of mobile phones stands out as exceptional. P2 mentioned that in the digital media era, smartphones act as an "external organ" for everyone, especially for the BLV community.}
\begin{quote}
\textit{\aaa{"Compared to sighted individuals, we rely on smartphones to meet basic life needs. Without them, many daily tasks (such as payment or navigation) would be impossible." }}\aaa{(P2)}
\end{quote}

\aaa{Without a smartphone, BLV users would face considerable inconvenience, even being unable to complete basic tasks. This indispensable function not only supports daily life but also enables BLV individuals to quickly and easily engage with games, highlighting the unparalleled convenience of mobile devices.}

\aaa{It is noteworthy that some prominent game companies have begun to design mobile games specifically for BLV users, leading the way for the BLV gaming community. P2, P7, P17, P22, and P24 all emphasized the importance of Chinese gaming companies dedicated to developing accessible mobile games. P18 is a media operator who mentioned that the company she works for specializes in developing mobile games and other internet social and entertainment products for the BLV community. Currently, the user base has surpassed 600,000. This supports the growing demand among BLV users for mobile games and other entertainment products. This attention has also increased the participation of BLV users in mobile gaming, fostering the development of broader online communities and more diverse social interactions.}

\subsection{Why Mobile Game?}

We found that on a personal level, the motivation of participants to play mobile games is not solely due to the high accessibility of smartphones. Instead, they turn to mobile games to alleviate boredom, gain a sense of accomplishment, and foster social connections.

\subsubsection{Alleviate Boredom}
\aaa{All our participants unanimously agreed that although some BLV individuals with higher education explore careers in sales, software testing, or product operations, the majority of BLV employment opportunities remain focused on traditional fields, such as traditional Chinese massage therapy and music tuning. }They argued that the limited employment opportunities often leave many of their BLV friends feeling trapped in repetitive cycles of monotonous work. Mobile games provide a valuable source of enjoyment for them and their friends. P16, a massage therapist, noted, 
\begin{quote}
    \textit{"My daily routine is massaging clients until late, which becomes monotonous. Mobile games, which are convenient and fun and I can play them during breaks at work or after my shift, bring me joy by allowing me to experience different careers and adventures, offering much needed relief from my exhausting job. "} (P16)
\end{quote}

Mobile games provide a unique form of convenience, allowing P16 to seamlessly transition from his work to a world of adventure. As our participants stated (N=6), some BLV individuals in China who lack opportunities to choose their preferred professions and are unwilling to work in massage therapy often remain unemployed after finishing high school or university. For them, mobile games offer a way to alleviate boredom and bring a sense of purpose. P15, a 25-year-old who has stayed home since graduating high school at 21, shared:
\begin{quote}
    \textit{ “The tactile pavement near my home is often blocked by shared bikes... For me, mobile games open up a virtual world where, even without leaving the house, I can explore new places, feel the freedom of flying or running in games... Different games allow me tome to take on roles like a doctor, warrior, or chef! Making my time on the phone feel meaningful and engaging.” } (P15)
\end{quote}

Therefore, mobile games provide a convenient way to relax and transition seamlessly from work to leisure. For most of the participants (N = 17), mobile games offer not only a means to alleviate boredom, but also a sense of purpose.

\subsubsection{A Sense of Achievement Difficult to Attain Elsewhere}

For many participants, video games create a space where they can feel competent and successful. The structured environment of a game allows them to pursue goals, overcome challenges, and receive immediate feedback, all of which contribute to a strong sense of accomplishment.

P16 had partial vision as a child and experienced many mainstream games that lacked any accessibility features for BLV players in his childhood. After gradually losing his sight at the age of 10, he continued to play different types of games based on his childhood memories and even helped congenitally BLV players enjoy gaming. P16 proudly stated, 
\begin{quote}
    \textit{"I play mobile games to feel joy and achieve a sense of accomplishment. Sometimes, that comes from social interactions; other times, it's about the thrill of winning—it feels amazing to win." }(P16)
\end{quote}

P10 recounted a memorable experience:
\begin{quote}
    \textit{\aaa{"When playing a strategic game, there was a difficult level with a very low official win rate. I initially couldn't pass it. After multiple attempts, I put in a lot of thought and calculated the optimal solution, and unexpectedly, I succeeded! The sense of achievement was simply incomparable."}} \aaa{(P10)}
\end{quote}

For some BLV players, mobile games have helped them release self-doubt and frustration that often build up over time. Engaging in gaming activities allows them to experience moments of success and belonging, which can be hard to find elsewhere. Many (N=14) reported feeling mentally healthier, noting that gaming had a therapeutic effect, offering them a much-needed break from the challenges of reality.

\subsubsection{A Pathway for BLV Players to Mainstream Inclusion}

Most BLV individuals in China grow up within the special education system, with limited opportunities to interact with students without disabilities, forming a distinct ecosystem exclusive to the BLV community. P19, a BLV mobile game player with twelve years of experience in China's special education system and four years at a U.S. university, mentioned that in her class in China, most students were blind or low vision, which meant they seldom relied on visual information and often discussed how sighted individuals might communicate about the same topics. P19 added, 
\begin{quote}
    \textit{"BLV people tend to be more direct and do not follow common social customs like handshakes or bows... Despite our strong desire to be part of mainstream society, there are few opportunities for us in China."} (P19)
\end{quote}

To BLV students, the special education system in China separates them from sighted people, making it difficult for them to interact with sighted peers. To adults, due to employment limitations, most people with disabilities work in environments where their colleagues also have similar disabilities, meaning that BLV individuals often interact primarily with other visually impaired coworkers. For many BLV players, mobile games have become a bridge connecting them to the broader world.

BLV players also aspire to engage with more mainstream games, as a way to obtain information from the external world. More than half of our participants (N=24) reported experiencing significant delays in receiving information. P19 leads an alliance in one of these BLV games and interacts with many BLV players. She explained her observations:
\begin{quote}
    \textit{“Sighted people can passively receive information at any time,but we have to actively search for information... So we often discover current trends much later! Back in school, I I loved reading sentences from youth pain literature. It was so popular at the time that social media platforms, driven by algorithms, kept recommending similar content to me. However, through discussions and interactions with others in games, I realized that my favorite style had been outdated for years. ”} (P19)
\end{quote}

\aaa{
Most social media platforms utilize algorithms to enable personalized content recommendations. However, for BLV individuals, such mechanisms based on limited historical preferences often result in outdated information. In contrast, games, as interactive media, provide BLV users with a fresh perspective on information access. Through real-time interactions within games, BLV players can transcend visual barriers, share information, and discuss topics with global players, capturing the latest trends in their areas of interest. Thus, games not only serve as information exchange platforms for the BLV community but also as effective channels for keeping up with the times and acquiring cutting-edge updates.}

\subsubsection{Fostering Deep Social Connections for BLV Users}
The BLV community has a deep desire for social connections, and gaming serves as one of the tools to develop these relationships. 

Mobile games help close the social distance between BLV individuals and sighted people by encouraging collaboration to complete tasks, while also providing common topics of conversation. For example, P4 enjoys playing mainstream mobile games like \textit{hide-and-seek} and \textit{parkour} with sighted players. These shared gaming experiences lead to conversations about mobile games in daily interactions and foster closer social relationships, enhancing her sense of inclusion in society. 

P19 described how she developed a close friendship while forming her alliance, explaining that she met her best friend seven years ago as teammates in a game alliance. They often played together, added each other on social media, and discovered a shared enjoyment in talking about life and hobbies, which led them to stay in touch and eventually become good friends over time.

In addition, gaming has become a breakthrough avenue for developing romantic relationships in the BLV community. Since offline social opportunities are limited, and they often don’t know where to find meaningful connections online, games serve as a platform for making these connections. P18, who works as a media operator, once promoted a BLV game by highlighting how it helped people find romantic partners: \textit{"Many people play this game because they heard that several couples met through it. I even posted some love letters from the game on our media promotion."} P27 also confirmed this through his gaming experience: 
\begin{quote}
    \textit{“I once experienced a couple hosting a grand traditional Chinese wedding in the game's main city. The groom rode ahead on horseback, while the bride followed in a sedan chair.”} (P27)
\end{quote}

This wedding left a lasting impression on the BLV player communities, and someone later shared an update, saying that the BLV couple had gotten together in real life as well.

\section{Findings: The Impact of Varying Levels of Accessibility in Mobile Games on BLV Players}\label{Findings:Accessibility}

According to previous literature and the consensus among our participants, mobile games accessible to BLV users can be categorized into three types based on the availability of accessible tools (with built-in voice reading systems of phones considered default features and not classified as accessible tools): (1) \aaa{BLV-specific mobile games;} (2) Mobile Games with Accessibility Tools; (3) Mobile Games Without Accessibility Tools. Most BLV players engage with mobile games specifically designed for the BLV, which typically have no visual elements and focus on sound effects, making them fully accessible for BLV users. Some mainstream mobile games in China fall into the middle category of "Mobile Games with Accessibility Tools", as they include limited accessibility tools to assist BLV players in exploring these games. However, the majority of games on the market, which are designed primarily for sighted players, lack accessibility optimizations and present significant barriers for BLV users.

\subsection{\aaa{BLV-Specific Mobile Games}} 
These games are specifically designed for BLV users and are referred to as "BLV games" within the BLV community. \aaa{All participants (N=32) mentioned that they had not encountered operational difficulties in these games. BLV games are equipped with built-in focus systems that simulate human vision by capturing screen information. Players can select different screen readers to audibly read this information aloud. The menu options and other settings are designed strictly following the habits of BLV users, adopting a list-style arrangement to prevent users from getting lost in complex menus.}

\aaa{Additionally, BLV games include comprehensive gesture systems, such as double-tap to confirm, swipe left or right to switch options, and semi-circle gestures to exit. China's mainstream BLV games have generally adopted a standardized operational model, which has become a universal convention that provides players with a preliminary understanding of game operations. Specifically, this model begins by determining the game’s orientation (portrait or landscape) and uses clear sound channels to indicate directions, laying the foundation for basic operations. The players then rely on these fundamentals to continue with the game.}

\begin{figure*}
    \centering
    \includegraphics[width=\linewidth]{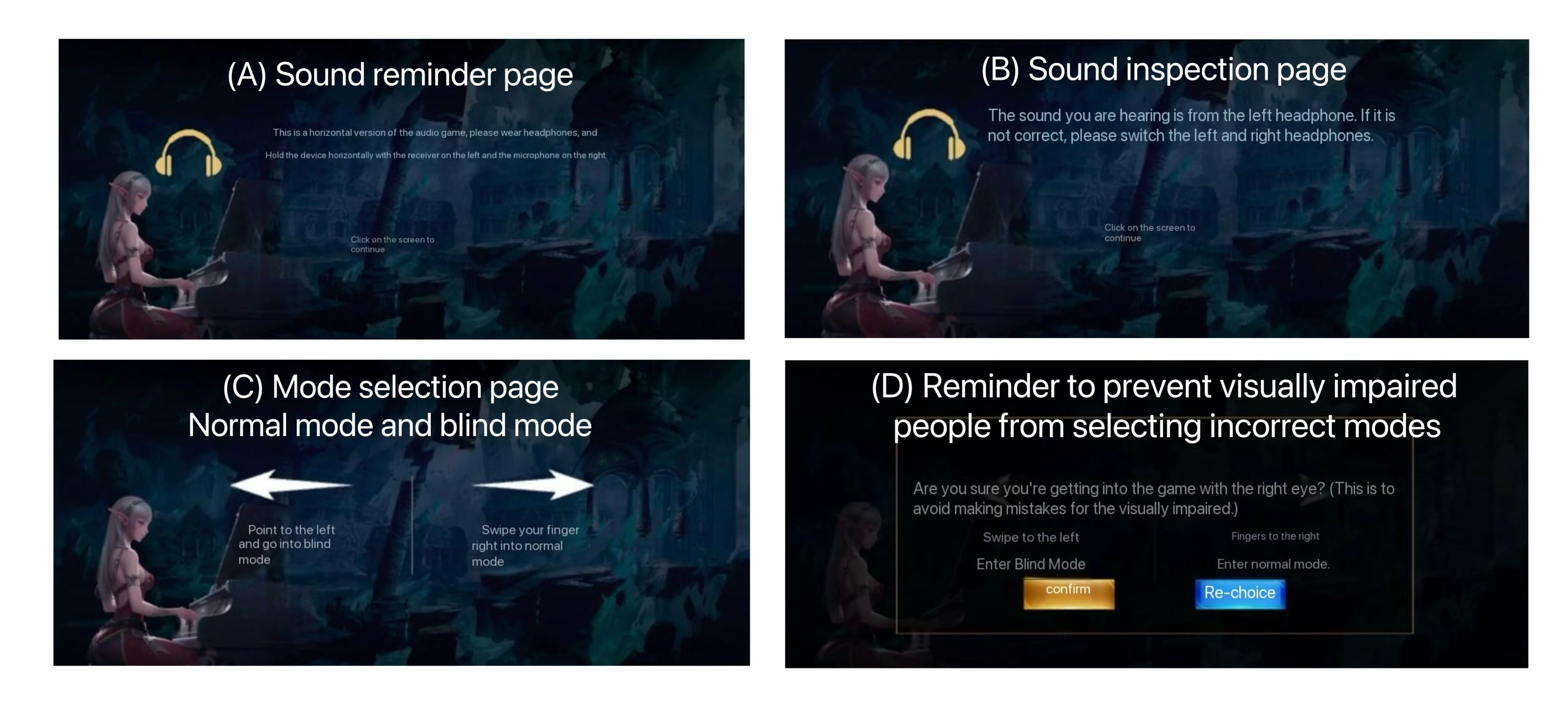}
    \caption{Four screenshots illustrate different interface pages for a BLV game, including sound reminder, sound inspection, mode selection, and a confirmation screen for visually impaired users to prevent incorrect mode selection.}
    \label{fig:enter-label}
\end{figure*}

\subsubsection{Types of 'BLV Mobile Games'}

Our interviewees (N=27) noted that, in China, card games, as the earliest type of BLV games, remain the most common form of gaming for the BLV community. BLV card game designer P1 mentioned that the cost of card game design is relatively low and offers greater flexibility in integrating accessible features.

\begin{quote}
    \textit{"Card games, with their simple and clear rules, align with the thought processes of BLV users and are also more strategic, allowing players to strategize and plan. Hence, our company focuses on developing turn-based card games."} (P1)
\end{quote}

According to BLV players we interviewed, in the past two years, some of the emerging BLV mobile games have been inspired by mainstream games designed for sighted players. This kind is different from the previously mentioned BLV mobile card games. According to P15, these BLV games are essentially adapted versions of popular mainstream games, retaining similar game mechanics and features while making adjustments for accessibility. They tend to offer high playability and have well-developed social systems, including official communities for sharing strategies. 

P15 mentioned a recently popular BLV game that is almost a copy of Genshin Impact, featuring a vast open world, real-time combat, and MMO elements such as world, guild, and team chat channels. This greatly enriched his gaming experience. \textit{"I play almost every day. The game is both accessible and offers a unique kind of fun compared to what I’ve experienced before. Although it took me a while to fully grasp how to play, I feel it was absolutely worth the effort."} However, as P15 pointed out, this kind of emerging BLV mobile games, which are designed to be similar to mainstream games, often have a higher operational threshold due to their extensive features and large scale. These emerging BLV mobile games involve many complex gestures and require extensive exploration. For designers, creating such games is also quite challenging. 

For congenital BLV users, opportunities to experience mainstream games designed for sighted players are extremely limited. Many of them are only familiar with BLV-specific games, making them the most loyal audience for these games. As P9 noted, many BLV players he knows have never played other types of games and, therefore, cannot make comparisons. This exclusivity fosters a high level of loyalty and satisfaction toward BLV games.

\subsubsection{Game Experience in 'BLV Mobile Games'}

When playing these BLV games, BLV players primarily seek to integrate into the BLV community. They often discover these games through recommendations from BLV game streamers or mutual suggestions within BLV player groups. Sharing similar gaming experiences helps them connect with more BLV friends. 

As P13 noted, \textit{"Since 99\% of users in the BLV gaming community are visually impaired, I feel this is a very safe space for conversation. Unlike interacting with sighted individuals, here we can freely joke about our visual impairments without fearing judgment."} 

P21 even met his wife while playing a BLV game: 
\begin{quote}
    \textit{"For a long time, my friends and I felt that our chances of entering marriage were slim... But my story has given many of my game partners hope. Therefore, BLV games tend to favor those who are more confident; perhaps games can help them develop healthy marital relationships."}(P21)
\end{quote}

For BLV games, design teams need to invest more effort in music and sound effects compared to games designed for sighted users. P31, who is very sensitive to sound, mentioned that he often does nothing but listen to the sound design in games. He excitedly shared his gaming experience: 
\begin{quote}
    \textit{"I prefer games feature 3D surround sound, background music for different maps - some areas have cheerful music, while mountain locations have more ethereal and clear music. These sound effects effectively convey information to us as players. For example, in a temple, we hear birds chirping, wind blowing, and chanting, signaling that 'we should find a monk to talk to'!"} (P31)
\end{quote}

\aaa{As a BLV game developer, P1 repeatedly emphasized the importance of sound design, many BLV games use heartbeat rates and breathing sounds in combat to express a character’s state, from healthy to lightly injured, to critically injured and near death, reflecting the intensity of the battle. This approach removes the cold, numerical voiceovers and allows players to use their imagination.}

\aaa{However, P7, as a game tester, emphasized that sound design needs to focus on the effectiveness of communication. When there is too much auditory information, the gaming experience can be diminished.}
\begin{quote}
    \textit{ \aaa{"In combat scenarios, players need to simultaneously monitor critical information such as skill announcements, positions, and health status. When multiple audio channels overlap, this can lead to information overload. Therefore, game design must differentiate between major and minor audio dimensions, prioritize sounds according to their relevance in specific contexts, and ensure seamless integration with other information."}} \aaa{(P7)}
\end{quote}

\aaa{Meanwhile, BLV games are beginning to explore more diverse uses of sound. For example, in a game developed by P1, there is a feature called "Lion's Roar," where the volume of the player's microphone input determines the strength of an attack on NPCs within the game. This two-way interaction with sound fully leverages the strengths of the BLV players.}

\aaa{Additionally, In terms of interface interaction, take selection boxes as an example. In traditional designs, elements such as watch-style scroll wheels use visual cues, highlighting the centered option while dimming surrounding choices, to indicate the linear scrolling direction. However, for BLV players, such visual cues are not intuitive. To optimize this experience, BLV games play specific sound effects when the screen reader focuses on the selection area. These sounds are designed to be both descriptive and directive, essentially saying,} \textit{\aaa{"This is where you need to scroll."}} \aaa{This auditory feedback not only lowers the usability threshold but also helps BLV players easily understand and complete the selection process. }

\subsubsection{Conflict Between Accessibility and Playability}

However, BLV games also have some shortcomings for BLV users. As previous research has noted, there is often a conflict between game accessibility and playability~\cite{bolesnikov2022understanding}. More than half of BLV players (N=18) have mentioned that BLV games are not very enjoyable, being overly simple and boring. \aaa{These games fail to satisfy the original motivations of BLV players for playing mobile games: boredom relief and a sense of achievement. A significant number of participants (N=5) mentioned that the activities in BLV games were too repetitive. }\textit{"We often predict what the next event will be in the group, and we always guess it correctly because it’s too simple. Eventually, we all quit playing."} (P11) 

P20, who was once an avid BLV player, has now uninstalled all BLV games. Regarding these games, P30 initially had high expectations, but eventually uninstalled all games due to the lack of achievement mechanisms, stating, 
\begin{quote}
   \textit{ "BLV games treat BLV players like idiots, offering extremely simple games. They seem to think that disability equals inability and that we cannot participate in more enjoyable games." }(P30)
\end{quote}

P13 tried both mainstream card games and card games designed for BLV groups. The former requires players to choose which card to play and when, while the latter just lets the system automatically play the cards once you enter the game.
\begin{quote}
\textit{"losing the strategic and intellectual elements that card games should have and does not treat us as players on the same level as sighted people."}(P13)
\end{quote} 

\aaa{In summary, universal mobile games suitable for BLV players primarily include card games and emerging mobile games influenced by mainstream titles. In these games, BLV players can seek close community connections and experience effective auditory communication. However, the lack of playability fails to satisfy the original gaming motivations of BLV players, and may even lead to ableism. BLV players may feel their abilities are being questioned, and their gaming experience could appear "inferior" or "unequal" in social comparison, causing them to lose confidence in gaming and potentially affecting their self-esteem.}

\subsection{Mobile Games with Accessibility Tools} 

Games with accessibility tools have specialized download channels. A very small portion is \textbf{ developed officially by companies}, while most are \textbf{created by BLV users themselves}. For example, P29 mentioned that a BLV player once designed a plugin for a multi-user domain game that captures information from scrolling text into a dialogue box to prevent missing information due to fast scrolling. There are also cases where BLV players have created plugins, and the official companies have tacitly accepted these plugins. For example, Blizzard's Hearthstone has allowed custom accessibility plugins created by BLV players. Among the interviewees (N=10), 4 have used officially developed plugins, while 8 have used plugins created by other BLV individuals.

\begin{figure*}
    \centering
    \includegraphics[width=\linewidth]{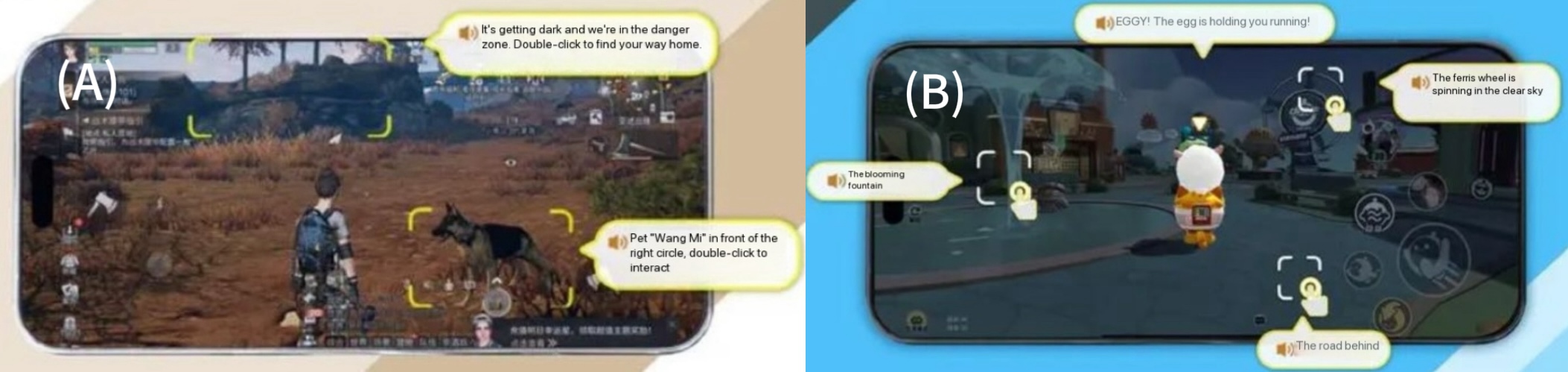}
    \caption{Games with a built-in AI voice assistant (AI EYES Mode): (A) The AI voice narrates the position of the character’s pet and the relative locations of surrounding mountains and buildings. (B) The AI voice narrates that the character is currently being lifted, and tapping around the screen will automatically describe objects and their positions.~\cite{AIEYES2024}}
    \label{fig:enter-label}
\end{figure*}

\subsubsection{Accessibility Tool and Game Experience}

BLV players often have high hopes for games designed for sighted players that include accessibility tools. Positioned between the most engaging and the most accessible games, these games with accompanying accessibility tools strike a relatively balanced position. They cater to delayed information reception for BLV players, offering an optimal blend of enhanced gameplay experience and smooth exploration. As P19 remarked, \textit{"BLV players cannot immediately adapt to sighted games without any tools. Games that fall into this middle threshold can mitigate our information delays and gently break down barriers between us and mainstream society."} 

Due to the time required to explore these games, even with accessible tools, those who persist are often individuals with a strong gaming background and a deep curiosity about visual games. In other words, the accessibility of these games is not easily achieved; instead, BLV players must exert significant effort, even in the stage of obtaining the necessary tools, long before they can truly begin playing. For example, P5, currently a student with ample free time, enjoys exploring these games on his own at school. Most participants (N=19) expressed a strong desire to use easy-to-use tools to explore mainstream games. P5 shared his heartfelt wish: 
\begin{quote}
   \textit{ “I firmly believe that even with my poor eyesight, if mainstream games were equipped with accessible features, I would be able to figure out how to play them. The most important thing is having the opportunity to try.”} (P5)
\end{quote}

As a streamer, P24 actively responded to the needs of BLV players, independently exploring coding, developing accessibility tools, designing accessible scripts, and creating video tutorials. He then shares these resources on BLV game forums or video platforms. To spread these evaluation videos to a wider BLV player group, he established a fan group, promoting the vibrancy and mobility of the BLV player community.
\begin{quote}
    \textit{"After finding and evaluating relevant accessible tools, I will share review videos of the plugins that I find suitable and useful. I will post them in communities and on social platforms to provide more information for BLV players, hoping that more people will be able to enjoy these games."} (P24)
\end{quote}

In these games, accessible tools typically provide a third-person perspective, describing ongoing events or assisting players in completing specific actions. For instance, according to P21, P23, and P32, with the right tools, \textit{Hearthstone} can capture every piece of information related to the game and competition, converting it into audio form to fill the visual gaps for low-vision players.

\subsubsection{Game challenges}
Similarly, these games also present certain challenges for BLV players.

First, installing plugins requires a certain level of technical proficiency. Most BLV players must search for tutorials through gaming communities or seek help from knowledgeable BLV players. Since the number of players willing to explore such games is relatively small, those enthusiastic about these games need to integrate into specialized gaming circles, communities with high levels of information and gaming literacy, to better learn how to play the game as a BLV person, exchange tips for clearing levels, or form teams for cooperative play. 

P2 and P8 indicated that without such information channels, it is difficult for BLV players to correctly download the accessibility tools. \textit{"Exploring accessibility tools on my own is too costly. Even after spending a lot of time, I might still get stuck in the first step, not knowing where to click or how to operate it,"} (P2)\textit{"These specialized gaming circles do not have public or official ways to join; membership usually happens somewhat by chance."} (P8) 

Meanwhile, the absence of detailed tutorials or fundamental programming knowledge makes the installation of auxiliary plugins challenging. P23 shared, \textit{"The accessibility tool was written in E-language, and my device kept giving error messages. I didn’t know what to do, so I ended up giving up."}

\aaa{Three participants shared similar experiences to P23. When downloading accessible tools, their devices would often emit error messages or security warnings. Sometimes, the mobile phone would freeze, crash or restart.} \aaa{P30 believes that, like BLV games, many accessibility tools for mainstream games could also be malicious software, increasing the risks they face when playing mainstream games. P22 mentioned that, due to the fear of such risks, he has always had a sense of apprehension towards mainstream games.}

Second, some games are not originally designed for the BLV community, and these plugins often function as workarounds rather than official features. They may not be reliable over time. After version updates or system patches, some plugins may become incompatible or be deemed as violations, exposing users to legal risks due to unauthorized software.

As a game tester, P7 is willing to experiment with various accessibility tools to explore this type of game, 
\begin{quote}
    \textit{"Since many games rely on visual information, these tools directly interpret that information and perform actions on our behalf, tasks that would typically require effort from sighted players... It feels like using 'cheat codes,' which is why these accessibility tools are often flagged as violations and usually become unusable."}(P7)
\end{quote}

Similarly, P4, P6, and P12 pointed out the "fragility" of these tools: \textit{"They might work one moment and become incompatible the next."} (P4)

Third, there is a risk of deception in the sharing of game installation packages or auxiliary plugins. Given the technical or informational barriers to access, some players exploit this information gap for profit through resource exchange. P27 shared, \textit{"In BLV gaming forums, people often message me privately, asking if I want to pay for accessibility tools to experience more games. It used to be that everyone shared these tools voluntarily, but I'm not sure when or why that changed."} P17 and P25, who have had relevant experience, noted that after making payments, they often receive incorrect toolsets, and sellers typically do not offer any further support.

\aaa{Games equipped with accessible tools strike a balance between playability and inclusivity. Compared to mainstream games, they offer higher accessibility while maintaining greater playability than many BLV-specific games. However, due to certain barriers in assistive functionalities, BLV players often expend additional effort and labor to participate in these games. This process may involve learning complex assistive features, customizing game settings, or seeking external guidance. These attempts often result in additional costs, both cognitive and emotional, which can undermine their confidence and willingness to engage with games.}

\subsection{Mainstream Mobile Games without Any Accessibility Tools}
Mainstream mobile games without accessibility tools have the largest market share and largest user base in China. Among our BLV interviewees, many (N=11) have tried these mobile games without accessibility tools, but most only gave them a cursory attempt. Only four BLV players have persisted with such games. Most BLV players struggle right from the initial login screen. A small number of players continue to experiment and develop their own operational systems.

\begin{figure*}
    \centering
    \includegraphics[width=\linewidth]{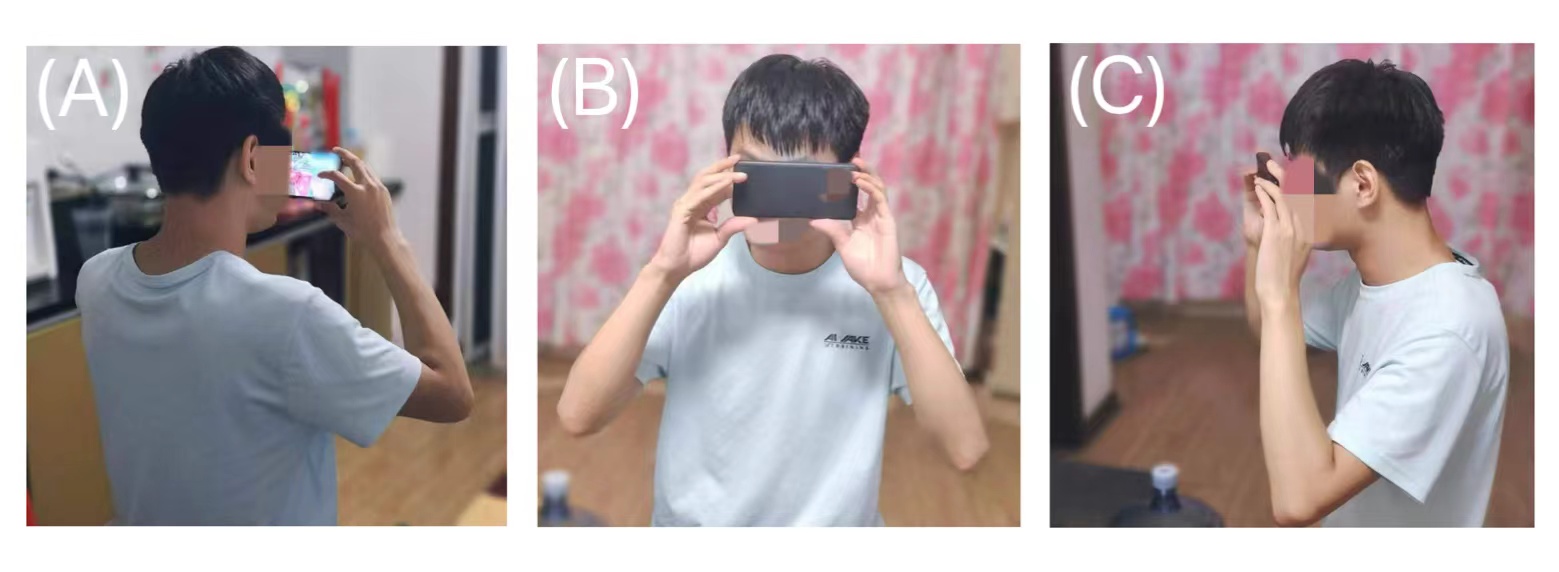}
    \caption{Our sampled BLV player is playing a mainstream mobile game designed for sighted players, holding the phone close to his face with both hands.
(A) The player, wearing a blue T-shirt, is shown from behind holding the phone at chest height, facing the screen closely.
(B) The player faces the camera, holding the phone directly in front of his eyes with both hands, positioning the screen close for visibility.
(C) A side view highlights the phone being held approximately 10 cm away from the player’s face for optimal viewing. The background depicts a casual indoor setting with floral wallpaper.}
    \label{fig:enter-label}
\end{figure*}

\subsubsection{Gaming Motivation and Experience}
\aaa{BLV players often express curiosity about the gaming experiences of the sighted mainstream society, they often grow bored with games designed specifically for them, criticizing the lack of diversity and challenges. As a result, they turn to mainstream games to seek richer gaming experiences.}

P32, who has tried various games, shared:

\begin{quote}
    \textit{"The daily task in the BLV game I play is to collect daily rewards and then start auto-battle, which is very boring. Occasionally, I see a streamer playing a mainstream game, coordinating hand movements, positioning, fighting monsters, and leveling up. It feels very immersive and interesting. From now on, I only want to play mainstream games."} (P32)
\end{quote}

Additionally, BLV players aim to use mainstream mobile games to strengthen their bonds with family, particularly younger family members. P2 noted that playing popular games allows him to stay updated on the latest trends, making it easier to connect with family and friends, thereby reducing feelings of isolation in a family setting. 

Moreover, some BLV players use these games as a means of interacting with sighted individuals. For certain BLV users, the main goal isn't necessarily gaming itself but rather initiating conversations with sighted players. For instance, P11 frequently initiates social interactions within these games:

\begin{quote}
    \textit{"I enjoy striking up conversations with people. I often learn about the latest trends and jokes through it. Without exaggeration, mainstream games have become my bridge to the outside world."}(P11)
\end{quote}

\aaa{As a gaming content creator, P24 regularly uploads videos on video-sharing platforms, showcasing his challenges of playing mainstream mobile games without any accessible tools.}

\begin{quote}
    \textit{\aaa{"I have over 7,000 followers, half of whom are BLV. I hope my videos help more people understand that BLV individuals can play games and play them well. At the same time, I want to show how challenging it is for BLV players to take on mainstream mobile games."}} \aaa{(P24)}
\end{quote}

\aaa{By sharing his experiences with mainstream mobile games, P24 not only enhances the BLV community's understanding of these games but also helps reduce the barriers to accessing them, thereby empowering other BLV players. The social interactions and video creation that stem from gaming contribute to increased visibility and recognition of BLV players, fostering better social integration and inclusion.}

\subsubsection{Barriers in Playing Mainstream Mobile Games without Any Accessibility Tools}

However, mainstream games without any accessibility tools present significant barriers for BLV players. Most sighted games rely heavily on visual information, with auditory elements limited to music or simple sound effects. \aaa{P18, a new media content operator, pointed out that the issue lies not in the visual information itself but in incomplete feedback:}

\begin{quote}
    \textit{\aaa{"For instance, when you enter a map and suddenly encounter a monster, BLV players might not see it, and without auditory cues, they can't position themselves properly. They might die unexpectedly without understanding why they failed or what to do next. This creates a terrible gaming experience."}} \aaa{(P18)}
\end{quote}

\aaa{Obstacles like these are meant to present fair challenges for all players, but without necessary sound cues for BLV players, they become "experience barriers." }

\aaa{BLV players require systems to translate visual information into auditory prompts, such as 1) screen readers for menus and controls, 2) clever sound effects (e.g., rapid heartbeats indicating low health), 3) ambient soundscapes for scene construction (e.g., crows and wind hitting tombstones to portray a gloomy graveyard), and 4) 3D audio for spatial awareness (e.g., changes in breathing or footsteps).}

\aaa{Additionally, some games feature designs that are particularly challenging for low-vision players. In order to enhance the visual and artistic effects, many mainstream games have incorporated frequent flashes on characters,} \textit{\aaa{"Designs like this puts extra strain on my eyes, causing them to feel sore and even worsening my vision."}} \aaa{ (P27)}

\aaa{Moreover, when some BLV users attempt to play mainstream games designed specifically for sighted players, the lack of accessibility tools can exacerbate psychological challenges such as autism and social anxiety. P26, who had repeatedly tried such games, expressed her sadness after facing multiple setbacks:}

\begin{quote}
    \textit{\aaa{"The convenience of mobile phones often makes me feel that visual impairment is no longer a barrier... However, when trying mainstream games designed for sighted players, the numerous accessibility barriers make me feel like a cyber blind person."}} \aaa{(P26)}
\end{quote}

\aaa{Among our participants, four pessimistically stated that the BLV community is destined to remain an overlooked user group in China's mainstream gaming industry.} \textit{\aaa{"Just thinking about mainstream games now fills me with fear—the fear of not being able to play and the sense of loss that comes with it. If I'm destined to feel disappointed, I'd rather not even start."}} \aaa{(P28)}

Psychological barriers caused by the low accessibility levels of mainstream games are not uncommon. Most participants believe that mainstream games are unlikely to accommodate visual impairments. These games often face issues such as incompatibility with screen readers, difficulty in locating buttons, and lack of instructional guidance. During their attempts, participants often encounter the warning "<! Tag Undefined !>". P3 sadly said that every "undefined" message feels like a reminder of their blindness and inability to play the game.

\subsubsection{Alternative Strategies and Design Expectations}

For mobile games that completely lack accessibility tools, BLV players have also explored alternative ways to play, often involving enlisting the help of sighted individuals, such as family or friends. By leveraging each other’s strengths, this collaboration can result in a highly effective and enjoyable gaming experience. P10, who enjoys puzzle mobile games, shared, \textit{“I have my cousin describe the visuals to me while I act as the strategist, directing him on how to play.”} P27 prefers to take a more hands-on approach: \textit{“I find a sighted person to inform me about the situation, and then I handle the gameplay myself. It’s quite enjoyable that way.”} Both methods essentially involve BLV players working with sighted individuals to bridge the visual gap, receiving real-time information.  BLV players then use their superior auditory skills to interpret the information, allowing them to make quicker and more accurate decisions about how to continue playing.

\aaa{
Participants highlighted several expectations for future game design, spanning visual, auditory, tactile features, and mechanics. P12 emphasized integrating screen reader systems with focus navigation during development to improve accessibility and reduce exploration time, encouraging BLV players to engage with games. P19 suggested defaulting games to accessibility mode to address initial setup barriers for BLV players, while allowing sighted users to disable it easily. P20 called for customizable interface options, such as adjustable contrast and font size, to enhance usability and immersion for low-vision players. Similarly, P21 proposed customizable audio feedback systems to improve clarity and guidance through adjustable sound settings. P26 stressed the need for accessible tutorials tailored to BLV players, enabling quicker onboarding and reducing the learning curve. P31 advocated for developers to engage with the BLV community through open communication channels to gather feedback and refine designs. Additionally, P11, P12, and P13 discussed AI companion features, with P13 suggesting AI-generated NPCs for dynamic open-ended interactions, making games more emotionally engaging.}

\aaa{Overall, participants envisioned mainstream mobile games that cater to BLV players’ diverse needs, combining accessibility with inclusivity to foster greater engagement and ease of play.}

\aaa{BLV players’ motivation to engage with mainstream games stems from curiosity about their appeal and a desire to use gaming as a medium for social integration. However, poor accessibility imposes significant barriers, leading to self-doubt and concerns about exclusion. Despite these challenges, participants shared adaptive strategies and expressed optimism for future improvements that could make games more inclusive and empowering.}

\section{Discussion}\label{sec:discussion}

Given the growing demand for mobile games among BLV users and their exposure to mainstream games designed for sighted players, examining the perceptions of these marginalized users regarding game accessibility is crucial for the future design of inclusive mobile games. 
Our findings show that BLV users' desire for mobile games stems from their aspiration to integrate into the mainstream society, but they often feel isolated due to limited accessibility, especially in the Global South ~\cite{smith2013digital,almog2018everyone}. For many BLV individuals, particularly those in the Global South, physical accessibility in their daily lives is severely limited, with insufficient employment and social opportunities~\cite{kameswaran2019experiences, palan2021seriously}. This lack of access drives them towards the digital society, especially the mobile aspect, where mobile games provide a significant alternative source of self-satisfaction and social interaction. Through the asset-based approach~\cite{wong2021reflections,pei2019we,wong2020needs}, we propose the following questions to encourage HCI scholars to reflect on how to enhance the game experience of users with disabilities by leveraging existing resources in situations where accessibility tools are limited:

\subsection{Gaming Practices and Resourcefulness Among BLV Individuals}

In line with previous studies~\cite{kameswaran2019experiences, palan2021seriously, chi2021franklin}, due to physical and social limitations, BLV individuals in countries with limited accessibility support often have restricted access to locations and social interactions, particularly with sighted individuals. In countries like China, for instance, BLV individuals may experience a segregated education system from an early age and enter distinctly different employment environments as adults~\cite{deng2007local,chen2016perspectives}. This segregation exacerbates their sense of alienation and intensifies their desire to integrate into mainstream society. In light of these challenges, other types of games also exhibit limited accessibility due to real-world constraints~\cite{coronado2023game}. For instance, computer games typically require specific hardware and a certain level of computer proficiency. Previous research has developed numerous accessibility tools for computer games~\cite{atkinson2006making,dobosz2016control,islam2020design,martinez2024playing,milne2014brailleplay}. However, challenges such as difficulties in operating computers and concerns about malicious software during downloads hinder their usability. Similarly, tabletop games generally require physical interaction, which is particularly challenging in the Global South, where offline environments for such activities are scarce, and this limitation further strains BLV users' social networks. 

\aaa{We demonstrate that the popularity of mobile games among the BLV community in China is driven by the widespread adoption of smartphones, the lack of computer skills and equipment, and the limited physical accessibility. In the absence of accessible facilities, we found that BLV individuals do not stop at these limitations but instead leverage their existing resources to engage in gaming. For instance, when they lack a computer and relevant skills, they do not pursue additional expenses to buy a computer or invest significant effort in learning new skills. Instead, they make use of their existing mobile devices and mobile games designed for them to seek gaming experiences. They also actively search for available accessibility tools within the information society to enhance their gaming experiences. Even in mainstream games that lack any accessibility tools, they explore their own experiences or seek alternative solutions, such as finding sighted players as gaming companions. This demonstrates the strong autonomy of BLV individuals in making the most of their existing assets.}

\subsection{Balancing Accessibility and Playability in BLV Gaming}

BLV games are designed to promote social interaction but face a conflict between playability and accessibility. Purely audio-based games are boring, while video games lack sufficient playability. Both academia and industry have developed accessible tools, such as AI Eyes, but these still have barriers and limited game support. BLV users are eager to engage in mainstream games, but without accessible tools, accessibility is low. They often learn about these games through videos but find it difficult to experience them firsthand, highlighting a strong demand for high-playability mainstream games with accessibility support.

\aaa{From an asset-based design perspective, we observed that BLV individuals do not passively accept the lack of accessibility options but instead actively leverage available resources to engage in gaming~\cite{rodrigues2020open,Yieke2024,yuan2008blind}. For example, when lacking computers or the necessary technical skills, they do not seek additional financial investments to purchase equipment or dedicate extensive effort to learn new skills. Instead, they utilize existing mobile devices and games specifically designed for the BLV community to access gaming experiences. They also actively seek accessible tools available within the information society to enhance their gameplay. Even in mainstream games without any assistive features, they find ways to engage by exploring the games themselves or collaborating with sighted players as gaming partners. }

This demonstrates the strong sense of agency among BLV individuals in maximizing the utility of available assets. Whether or not these games include accessibility tools, the act of playing them represents an effort to bridge the gap between their own experiences and those of sighted players.

\subsection{Addressing Ableism and Cyber Blindness in Mobile Game Design}

\aaa{Our research reveals that ableism is deeply ingrained in mobile game design, prioritizing sighted users while marginalizing BLV players. Although audio-based features and assistive technologies have been developed, they often fail to address systemic issues like visibility management and ethical inclusivity. Mainstream games heavily depend on visual elements, e.g., complex graphics, dynamic effects, and visual environments, rendering them inaccessible to BLV players and exacerbating their social exclusion. Even games specifically designed for BLV players primarily focus on auditory feedback while neglecting critical visual adjustments like customizable color contrast and font size.}

\aaa{Cairns et al. mentioned that if game design is not done properly, it can actually rebuild disability in the digital realm and even create new, unexpected barriers~\cite{cairns2021enabled}. In mobile games lacking proper accessibility tools, technological limitations prevent BLV players from enjoying the same gaming experience as sighted players, and their physical limitations are still forcefully exposed. This reflects a broader issue of "Cyber Blindness," where technological flaws amplify the social manifestations of disability, leading to emotional harm and challenging BLV players' sense of identity. Mainstream games, which lack accessibility tools, make it difficult for BLV players to experience the same gameplay as sighted players, while BLV games are limited to the BLV gaming community. When BLV individuals attempt to integrate into mainstream gaming communities through technology, the technical barriers they face often make them feel marginalized or excluded in the experience, further reinforcing the gap between them and sighted players.}

\aaa{To address these challenges, game design must transcend accessibility to embrace "visibility control," allowing BLV players to manage how their disability is perceived during gameplay. A promising approach involves introducing dual exploration modes, such as "Sighted Mode" and "Visually Impaired Mode," rather than creating separate versions of games. This approach fosters inclusivity, enabling BLV players to engage with mainstream games on equal terms while maintaining autonomy and dignity. By eliminating misconceptions and reducing the potential for harm, designers can create games that respect all players’ identities, fostering a more inclusive and ethical gaming environment where everyone can participate and feel valued.}

\section{Limitations}
Firstly, the study focuses mainly on BLV users in China. Cultural and regional differences in technology use, accessibility standards, and social norms may influence the applicability of the results to BLV communities in other countries or regions. Secondly, although this study involved four BLV game designers and offered valuable insights into BLV game design, it lacks a technical and commercial perspective, particularly from those involved in designing mainstream games. The research focuses primarily on the experiences of BLV users and their feedback on existing games, with limited analysis of the design processes, decisions, and how mainstream game designers consider or overlook accessibility factors. This gap means that we may not fully understand the challenges and opportunities present in mainstream game development, nor how designers balance commercial goals with accessibility needs. Finally, this study primarily focuses on the experiences and needs of BLV users and does not thoroughly explore the gaming needs of individuals with other types of disabilities. Users with other disabilities, such as those with mobility impairments, hearing loss, or cognitive disabilities, also constitute an important segment of the gaming community. Each disability presents unique challenges and requirements for game accessibility, and understanding these needs is crucial for creating inclusive gaming experiences. 

\section{Conclusion}
In conclusion, our study sheds light on the experiences and motivations of BLV players in the mobile gaming landscape. Our study found that BLV players are highly motivated to engage with mobile games as a means of escaping real-world limitations, achieving self-satisfaction, and fostering social connections. Mobile games, with their inherent accessibility features, are a more viable option compared to tabletop and computer games. We explored the impact of varying levels of game accessibility on BLV players and found that accessibility significantly influences their gaming experience. 
Overall, our findings highlight the importance of narrowing the accessibility gap in games and \aaa{suggest that future research should focus on the psychological and experiential barriers faced by BLV players in mainstream mobile games due to technological inequality. Specifically, there should be more exploration into the relationship between sound and visual communication effectiveness, game mechanics design, playability, and accessibility. }Additionally, we recommend that future studies explore methods to enhance the gaming experience for BLV and other users with different disabilities, enabling them to independently explore mainstream mobile games.

\bibliographystyle{ACM-Reference-Format}
\bibliography{sample-base.bib}

\appendix 

\end{document}